\documentclass[conference]{IEEEtran}
\IEEEoverridecommandlockouts
\usepackage{cite}
\usepackage{amsmath,amssymb,amsfonts}
\usepackage{algorithmic}
\usepackage{graphicx}
\usepackage{textcomp}
\usepackage{xcolor}
\usepackage{float}
\usepackage{booktabs}

\usepackage{acronym}
\usepackage{subcaption}
\usepackage{float}
\usepackage{multirow}
\usepackage{soul}
\usepackage{comment}

\usepackage{enumitem,amssymb}
\newlist{todolist}{itemize}{2}
\setlist[todolist]{label=$\square$}
\usepackage{pifont}
\usepackage{array}
\newcolumntype{P}[1]{>{\centering\arraybackslash}p{#1}}
\newcolumntype{M}[1]{>{\centering\arraybackslash}m{#1}}

\def\BibTeX{{\rm B\kern-.05em{\sc i\kern-.025em b}\kern-.08em
    T\kern-.1667em\lower.7ex\hbox{E}\kern-.125emX}}
\begin{document}

\acrodef{FR}{frame rate}
\acrodef{WULPUS}{Wearable Ultra Low-Power Ultrasound}
\acrodef{US}{Ultrasound}
\acrodef{ULP}{ultra low-power}
\acrodef{AFE}{analog front-end}
\acrodef{RF}{radio frequency}
\acrodef{ML}{machine Learning}
\acrodef{FPGA}{Field-Programmable Gate Array}
\acrodef{LVDS}{low-voltage differential signalling}
\acrodef{SPI}{serial peripheral interface}
\acrodef{MCU}{microcontroller unit}
\acrodef{ADC}{analog to digital Converter}
\acrodef{HDL}{hardware description language}
\acrodef{DMA}{direct memory access}

\acrodef{HD}{hardware design}
\acrodef{AD}{algorithm development}
\acrodef{FPS}{frames per second}
\acrodef{DAS}{delay-and-sum}
\acrodef{TGC}{time-gain compensation}

\acrodef{PULP}{Parallel Ultra Low Power}
\acrodef{SoC}{System on Chip}
\acrodef{BLE}{Bluetooth Low Energy}
\acrodef{EEG}{electroencephalography}
\acrodef{EMG}{electromyography}
\acrodef{ECG}{electrocardiogram}

\title{Open-Source Multi-Wire SPI Readout for Wearable Ultrasound Probes

\thanks{The authors acknowledge support from the ETH Research Grant ETH-C-01-21-2 (Project ListenToLight) and from the ETH Future Computing Laboratory (EFCL).}
}

\author{
  \IEEEauthorblockN{
    Federico~Villani\IEEEauthorrefmark{1},
    Soumyo~Bhattacharjee\IEEEauthorrefmark{1},
    Lisa~Odermatt\IEEEauthorrefmark{1},
    Cédric~Hirschi\IEEEauthorrefmark{1},
    Luca~Benini\IEEEauthorrefmark{1}\IEEEauthorrefmark{2},
    Andrea~Cossettini\IEEEauthorrefmark{1}\\}
    
    \vspace{0.2cm}

  \IEEEauthorblockA{\IEEEauthorrefmark{1}Integrated Systems Laboratory, ETH Z{\"u}rich, Z{\"u}rich, Switzerland}
  \IEEEauthorblockA{\IEEEauthorrefmark{2}DEI, University of Bologna, Bologna, Italy}
}

\maketitle

\begin{abstract}
Wearable ultrasound probes must transfer increasingly large acquisition payloads while maintaining compact, low-power electronics. In the 32-channel TinyProbe platform, this transfer is bounded by the single-lane SPI interface between the acquisition FPGA and the wireless system controller. This work presents an open-source, multi-lane SPI readout interface that uses serial command and address phases followed by a build-time-selectable dual- or quad-lane payload phase, thereby increasing throughput at an unchanged clock rate. 
The interface emulates a serial flash memory, enabling compatibility with a broad range of microcontroller families and their existing peripheral interfaces. On the FPGA, the data path connects the existing acquisition FIFOs to the SPI interface through clock-domain crossing, sample reshaping, and packing into 32-bit words. 
A Kria K26 testbed is used to characterize the FPGA SPI interface independently of the acquisition and wireless subsystems, demonstrating error-free transfers at SCLK frequencies up to 66 MHz, i.e., 132~Mbit/s in dual-lane and 264~Mbit/s in quad-lane mode, 3.3$\times$ and 6.6$\times$ the 40 Mbit/s single-lane baseline.
Integrated in TinyProbe without hardware modification, dual-lane readout doubles the rate at the same SCLK and identifies the SiWG917 multi-lane SPI implementation as the next bandwidth-limiting component. 
\end{abstract}


\section{Introduction}
\label{sec:introduction}

Wearable ultrasound enables continuous, unhindered monitoring of deep tissue structures and physiological dynamics using compact, programmable probes \cite{zhou2025wearable}, with demonstrated applications across cardiovascular \cite{hu2023wearable,sonus} and internal organ moitoring \cite{toymus2024integrated,liu2024wearable}, musculoskeletal sensing\cite{ sgambato2024hmi,lu2026hand}, and other domains.

Multichannel wearable probes have enabled modalities that are beyond the capabilities of single-element or few-channel devices, such as beam steering, B-mode imaging, and plane-wave compounding.  Recent examples also demonstrate their successful use on moving subjects \cite{linFullyIntegratedWearable2024}. In this context, the 32-channel TinyProbe platform \cite{vostrikovTinyProbeWearable32Channel2025a} stands out for its high-speed capabilities (32-channel frame rates of 175 Hz, 2-ch. Pulsed Wave Doppler at 1.5 kHz), enabled by a Wi-Fi~6 link with a 35.7 Mbit/s throughput~\cite{hirschiHighFrameRate2025}.

This throughput is limited not by the radio interface but by the preceding FPGA-to-controller stage.
%
%
In fact, TinyProbe features a conventional SPI interface for data movement between the FPGA and the system controller, whose bandwidth constrains the pulse-repetition frequency and limits the range of experiments that can be performed without reducing the acquisition payload \cite{hirschiHighFrameRate2025}.

This work addresses that limitation at the FPGA side of the readout stage.
We present an open-source multi-lane SPI readout interface in which the command and address remain serial while the payload is transferred over two or four data lines.
The interface emulates a serial flash memory, and in dual-lane mode it reuses the two conductors already routed between FPGA and TinyProbe system controller, requiring no hardware modification of the probe. 

The main contributions of this work are:
\begin{itemize}
    \item a multi-wire SPI protocol and FPGA data path providing register access and buffered acquisition readout, with a build-time selectable dual- or quad-lane payload phase;
    \item its characterization up to 66~MHz SCLK (264~Mbit/s in quad-lane mode) on a Kria K26 FPGA testbed, demonstrating rates 3.3$\times$ and 6.6$\times$ the 40~Mbit/s single-lane link of \cite{hirschiHighFrameRate2025}, together with static timing closure of the TinyProbe IGLOO2 implementation;
    \item its integration and verification in the TinyProbe readout chain over the existing conductors, quantifying the remaining controller-side limitation, and defining the requirements for the next system controller
\end{itemize}

The HDL and MCU implementations are planned to be released under a permissive open-source license at \mbox{https://github.com/pulp-bio/TinyProbe}
\section{Materials and Methods}
\label{sec:architecture}

\subsection{TinyProbe Readout Architecture Baseline}

\begin{figure*}[t]
    \centering
    \includegraphics[width=\textwidth]{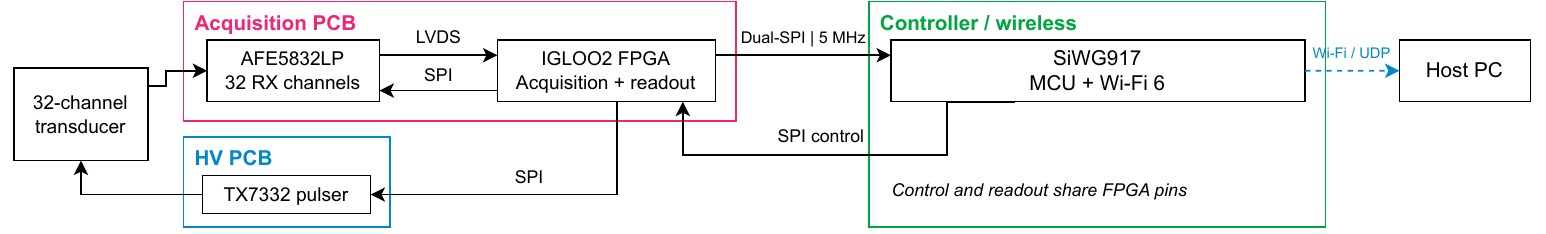}
    \caption{TinyProbe architecture with the proposed dual-SPI readout. }
    \label{fig:system_overview}
\end{figure*}

The interface presented in this work is designed for integration into the 32-channel Wi-Fi 6 version of the TinyProbe platform~\cite{vostrikovTinyProbeWearable32Channel2025a,hirschiHighFrameRate2025} (Fig.~\ref{fig:system_overview}). A TX7332\cite{texasinstrumentsTX7332ThreeLevel32Channel2019} pulser excites the transducer array, and an AFE5832LP\cite{texasinstrumentsAFE5832LP32ChannelUltrasound2018} analog front-end (AFE) amplifies and digitizes the received echoes, which are then transferred over 16 LVDS lanes to an IGLOO2 M2GL050-VFG400\cite{microchiptechnologyinc.IGLOO2M2GL050VFG400FPGA2026} FPGA. The FPGA generates the acquisition timing and buffers the samples in one FIFO per lane. A SiWG917 system-on-chip, which integrates a 2.4 GHz Wi-Fi 6 radio, 
configures the probe (FPGA, AFE, and pulser) and forwards acquisitions to the host over UDP at up to 35.7 Mbps \cite{hirschiHighFrameRate2025}.

The raw-data throughput from probe to host is therefore bounded by two serial stages: \textit{(i)} FPGA-to-controller readout and \textit{(ii)} controller-to-host wireless transmission. In this paper, we address the FPGA stage \textit{(i)}.

In the baseline, stage \textit{(i)} consists of a single full-duplex 40~MHz SPI link, with the SiWG917 as controller, shared by configuration and readout. Configuration of the pulser and AFE is performed through a target-selection and pass-through logic, in which the FPGA acts as a multiplexer of the full-duplex single-lane SPI. 
Both functions use the same data conductors (MOSI and MISO), and
because a single data lane cannot exceed SCLK, higher throughput requires adding additional lanes.

\subsection{Multi-Lane SPI Protocol}

The proposed FPGA core implements an SPI secondary operating on a subset of standard flash memory transactions to support a wide variety of microcontroller peripherals designed primarily for flash memory support.  Each transaction starts with an 8-bit opcode, either register write (\texttt{0x32}), register read (\texttt{0x4C}), or buffered stream read (\texttt{0x6B}), followed by a 24-bit address, both transferred serially on IO0, most-significant bit first. After the command and address phases, the payload is transferred over $L$ data lanes (with $L=2$ or $L=4$, fixed at synthesis). A 32-bit payload word requires $32/L$ SCLK cycles, i.e., 16 in dual-lane mode and 8 in quad-lane mode.

Read transactions include ten dummy SCLK cycles between the address and payload phases. This interval provides the turnaround time for register responses and buffered data. During FPGA-to-controller transfers, the output drivers are enabled only for the active payload lines; outside the transmit-data phase, the FPGA data outputs are tri-stated. 

In dual-lane mode, the two lanes are the conductors used as MOSI and MISO by conventional SPI (to guarantee compatibility with the existing SiWG917 connection).
The two chip select pins implement a four-state multiplexer: (0,0) selects the SPI subsystem, (0,1) the AFE pass-through, (1,0) the TX pass-through, and (1,1) the internal register interface and stream read. In state 3, the command and address remain single-lane, while the payload phase switches to half-duplex dual-lane operation. In this mode, register write and register read are used to program the FPGA's register map, and buffered stream read is used to read acquired data to the SPI main. The use of these two modes ensures full use of the limited SPI data pins and doubles the potential data rate for the buffer read operation.



\subsection{FPGA Data Path}

\begin{figure*}[t]
    \centering
    \includegraphics[width=\textwidth]{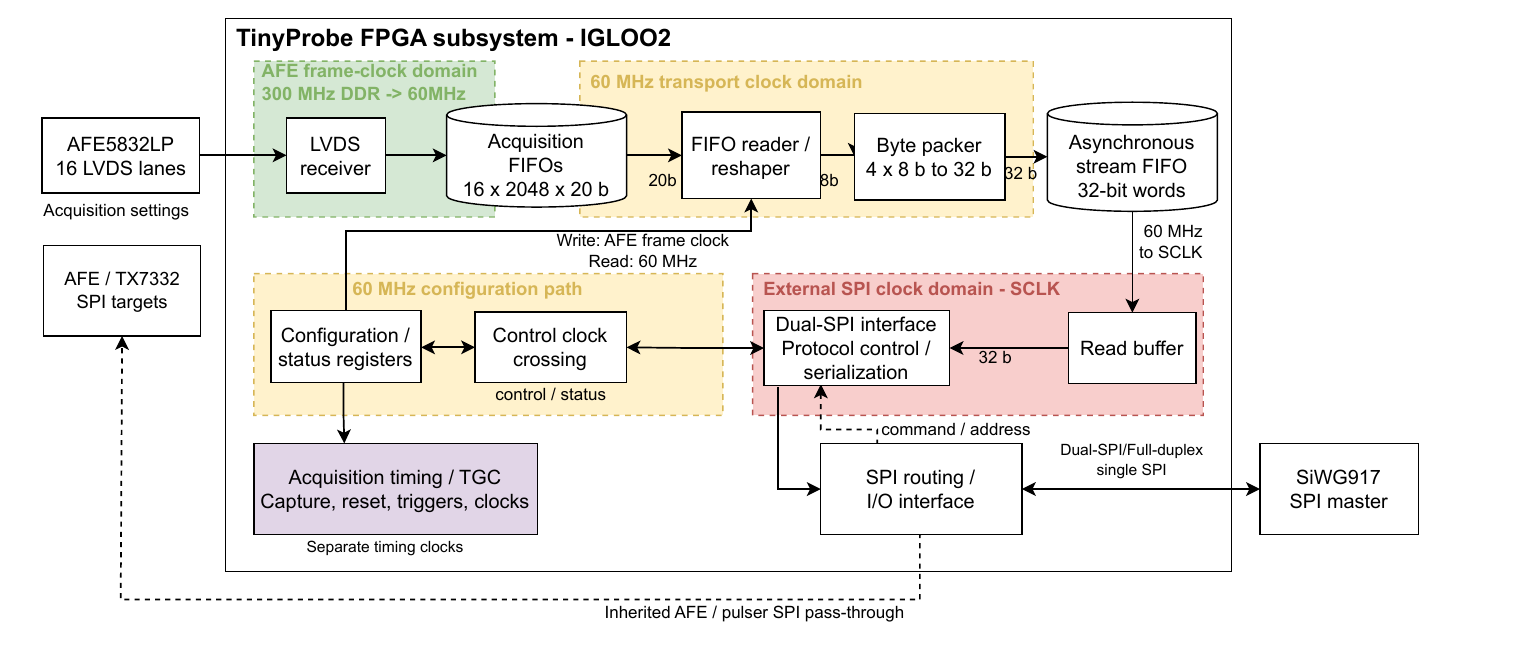}
    \caption{TinyProbe FPGA acquisition and dual-SPI readout data path. Control operations cross between the SPI and transport clock domains, while acquisition data are read from the TinyProbe FIFOs, reshaped and packed into 32-bit words, buffered asynchronously, and transferred through the multi-wire SPI interface.}
    \label{fig:readout_ip}
\end{figure*}

Fig.~\ref{fig:readout_ip} shows the FPGA data path, which separates low-rate control from burst acquisition data. 
Register writes and read requests cross from the SPI clock domain to the 60~MHz transport clock domain through request/acknowledge mailboxes. Read responses are returned to the SPI domain through the same clock-domain-crossing structure. This control path is connected to the existing 16$\times$32-bit TinyProbe register bank and its associated status and control logic.

Acquisition data follow a dedicated path through the FPGA. Each LVDS ADC lane carries two 10-bit channel samples that are written to the corresponding acquisition FIFO. Each FIFO entry contains two 10-bit values. An acquisition finite-state machine (FSM) controls the capture sequence and clears the FIFOs at the beginning of each acquisition. Once the acquisition is complete, the stored samples are ready to be transmitted to the MCU main.

The existing multi-FIFO reader traverses the enabled acquisition FIFOs and generates a FIFO-lane-major stream. Each FIFO corresponds to one AFE ADC lane serving a pair of receive channels. The native 20-bit words are reshaped into an 8-bit byte stream, after which four bytes are packed into each 32-bit transport word. An asynchronous FIFO transfers these words into the SPI clock domain. During a \texttt{0x6B} read transaction, complete 32-bit words are supplied to the protocol engine and serialized over the selected dual- or quad-lane payload interface.

A standard TinyProbe acquisition contains 2048 20-bit words per active FIFO, corresponding to 5120 bytes, or 1280 32-bit transport words, per FIFO. With all 16 FIFOs enabled, this results in 81,920 bytes per acquisition across 32 receive channels. This payload size is determined by the acquisition configuration and is not an intrinsic transaction-length constraint of the multi-wire SPI protocol.

\subsection{Characterization}
\label{sec:demonstration}

\begin{figure}[t]
    \centering
    \includegraphics[width=\columnwidth]{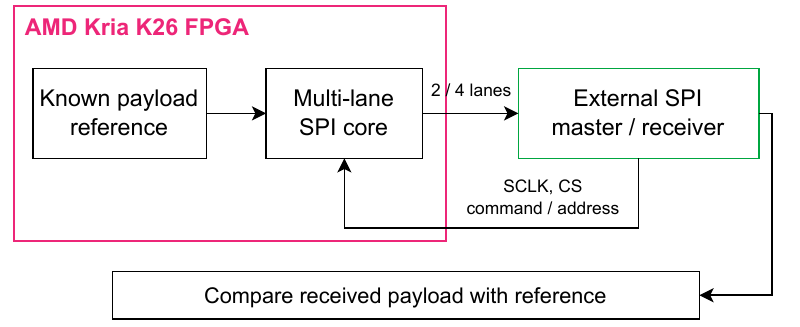}
    \caption{Kria K26 testbed used to characterize the proposed FPGA interface. A known FPGA payload is transferred to an external SPI main and compared with the reference sequence.}
    \label{fig:characterization}
\end{figure}

To characterize the FPGA interface independently from the integrated controller, the same protocol core is implemented on an AMD Kria K26. As shown in Fig.~\ref{fig:characterization}, a known payload generator supplies a ramp pattern, and an external STM32N657X0H3Q SPI main receives the dual- or quad-lane stream. The received payload is compared with the reference sequence while SCLK is varied. A configuration passes if no errors are detected on the payload.  This setup isolates protocol and FPGA I/O behavior from the TinyProbe acquisition reader, controller software, and wireless link.

Additionally, since the deployed FPGA differs from the testbed FPGA, we verify an IGLOO2 implementation via static timing analysis at SCLK~=50~MHz.

\subsection{Integration into TinyProbe}

Finally, we integrate the proposed solution in the complete TinyProbe stack.
The SiWG917 acts as SPI main and communicates with the IGLOO2 acquisition FPGA. The existing target-selection logic is retained for conventional FPGA, AFE, and pulser configuration. For acquisition readout, the controller selects the FPGA, issues the stream-read command, receives the dual-lane payload, and forwards the acquired data over the existing UDP bulk-data path to the host. 

In multi-lane mode, the SiWG917 limits SCLK to approximately 10~MHz, compared with 40~MHz in single-lane mode. \footnote{https://docs.silabs.com/wifi-developer-guides/1.0.3/peripherals-ssi/ssi-appendix\#q-what-is-the-maximum-serial-peripheral-interface-spi-clock-frequency-supported}
We therefore validate the integrated hardware at a conservative 5~MHz SCLK, which verifies end-to-end functionality on the unmodified TinyProbe.
Removing this controller-side limitation is left for future work (see Sect.~\ref{sec:discussion}).

One acquisition proceeds as follows: 
\begin{enumerate}
    \item The controller configures the AFE and pulser over conventional SPI through the FPGA pass-through and register interface. 
    \item The FPGA generates the acquisition triggers for the pulser and AFE, starting the acquisition, and storing the received samples in its internal FIFOs.
    \item The controller selects the FPGA and issues a \texttt{0x6B} stream-read command. This command changes the FPGA's conventional SPI interface to dual-SPI mode\footnote{During the dual-lane read payload phase, the MOSI conductor reverses direction so that both data lines carry FPGA-to-controller data.}, forming a two-lane payload phase. 
    \item The FPGA reads the acquired samples from the FIFOs, repacks them into 32-bit transport words, and transfers them to the controller over dual-SPI.
    \item The controller receives the payload via direct memory access (DMA) and transmits it to the host over UDP.
\end{enumerate}

\section{Results and Discussion}
\label{sec:results}

\subsection{Isolated FPGA Characterization}

The custom SPI block was verified on a Kria KR260 development board in order to measure the maximum operating frequency. Four data pins, one clock pin, and one chip select pin of the I/O bank 45 of the K26 module  were connected to the STM32N657X0H3Q XSPI1 peripheral. The XSPI1 speed was gradually increased until streamed reads of the internal buffer memory returned errors. No transaction errors were detected up to a speed of 66~MHz in quad mode, resulting in a maximum raw bitrate of 264~Mbit/s. At the same SCLK frequency, the corresponding raw bitrate for dual mode is 132~Mbit/s.

\textit{IGLOO2 Analysis.} The IGLOO2 implementation meets timing at 50~Mhz and is extremely resource efficient, requiring 187 LUT and 161 FF logic elements, each less than 1.25\% of the complete TinyProbe design resource usage.

\subsection{TinyProbe Integration}

The IGLOO2/SiWG917 demonstrator successfully completed acquisition-buffer readout and UDP forwarding at 5~MHz SCLK in dual-SPI mode. Two payload bits are transferred per SCLK cycle, corresponding to a calculated raw payload-phase rate of 10~Mbit/s. The experiment confirms functional integration with the existing TinyProbe conductors and readout software.




\subsection{Discussion and Future Work}
\label{sec:discussion}

The isolated FPGA characterization shows that the multi-lane interface can operate at substantially higher frequencies than those supported by the current TinyProbe controller. The FPGA side of stage \textit{(i)} is therefore no longer the limiting element of the readout path.

In the integrated system, however, the SiWG917 SPI host limits SCLK to approximately 10~MHz in multi-lane mode\footnote{https://docs.silabs.com/wifi-developer-guides/1.0.3/peripherals-ssi/ssi-appendix\#q-what-is-the-maximum-serial-peripheral-interface-spi-clock-frequency-supported}. Thus, with this specific MCU, widening the FPGA interface alone does not allow the system to exploit the demonstrated FPGA-side bandwidth.

This limitation motivated the proposed next-generation architecture. Fully exploiting a larger fraction of the demonstrated FPGA-side rate requires a controller providing multi-lane SPI operation at $\geq$~50~MHz and enough conductors to separate configuration from readout. Devices such as the STM32N6 meet these requirements, and future work will focus on replacing the SiWG917 with such an alternative paired with a dedicated wireless coprocessor.






\section{Conclusion}
\label{sec:conclusion}

We presented a multi-wire SPI readout interface for the TinyProbe wearable ultrasound probe, comprising a flash-style SPI peripheral with build-time selectable dual- or quad-lane payload phases while retaining serial command and address phases. The FPGA data path connects the existing acquisition FIFOs to the interface through sample reshaping, 32-bit packing, and asynchronous buffering.

On a Kria K26 testbed, the interface was verified up to 66~MHz in quad mode, corresponding to a raw payload-phase rate of 264~Mbit/s. The interface was further integrated into the 32-channel TinyProbe using the existing connection between the IGLOO2 FPGA and the SiWG917 controller and successfully demonstrated dual-SPI acquisition readout and UDP forwarding.

The integration identified the SiWG917 multi-lane SPI host as the remaining limitation of the readout path, motivating the proposed next-generation backend with a faster multi-lane SPI controller and dedicated wireless subsystem. We plan to open source the proposed FPGA subsystem at \mbox{https://github.com/pulp-bio/TinyProbe}.






\section*{Acknowledgment}

We thank Thomas Quanbrough (ETH Z{\"u}rich) for technical support.


\bibliographystyle{IEEEtran}
\bibliography{bibliography.bib}

\end{document}